\documentclass[aps, prl, reprint, superscriptaddress]{revtex4-1} 
\usepackage{amsmath,graphicx}
\usepackage[colorlinks, allcolors=blue]{hyperref}
\def\eu{\mathrm{e}}
\def\iu{\mathrm{i}}
\def\du{\mathrm{d}}

\begin{document}
\title{Chiral Magnetic Photocurrent in Dirac and Weyl Materials}
\date{\today}
\author{Sahal Kaushik}
\email{sahal.kaushik@stonybrook.edu}
\affiliation{Department of Physics and Astronomy, Stony Brook University, Stony Brook, New York 11794-3800, USA}
\author{Dmitri E. Kharzeev}
\email{dmitri.kharzeev@stonybrook.edu}
\email{kharzeev@bnl.gov}
\affiliation{Department of Physics and Astronomy, Stony Brook University, Stony Brook, New York 11794-3800, USA}
\affiliation{Department of Physics, Brookhaven National Laboratory, Upton, New York 11973-5000, USA}
\affiliation{RIKEN-BNL Research Center, Brookhaven National Laboratory, Upton, New York 11973-5000, USA}
\author{Evan John Philip}
\email{evan.philip@stonybrook.edu}
\affiliation{Department of Physics and Astronomy, Stony Brook University, Stony Brook, New York 11794-3800, USA}

\begin{abstract}
Circularly polarized light (CPL) can induce an asymmetry between the number of left- and right-handed chiral quasiparticles in Dirac and Weyl semimetals. 
We show that if the photoresponse of the material is dominated by chiral quasiparticles, the total chiral charge induced in the material by CPL can be evaluated in a model-independent way through the chiral anomaly. In the presence of an external magnetic field perpendicular to the incident CPL, this allows to predict the linear density of the induced photocurrent resulting from the chiral magnetic effect. The predicted effect should exist in any kind of Dirac or Weyl materials, with both symmetric and asymmetric band structure. An estimate of the resulting {\it{chiral magnetic photocurrent}} in a typical Dirac semimetal irradiated by an infrared laser of intensity~$\simeq 5 \times 10^6\, \mathrm{W/m^2}$ and a wavelength of $\lambda \simeq 10\, \mu\mathrm{m}$ in an external magnetic field $B \simeq 2\, \mathrm{T}$ yields a current  $J \simeq 50\,\mathrm{nA}$ in the laser spot of size~$\simeq 50\,\mu\mathrm{m}$. This current scales linearly with the magnetic field and wavelength, opening up possibilities for applications in photonics, optoelectronics, and THz sensing.
\end{abstract}
\maketitle

\section{Introduction}
Circularly polarized light (CPL) breaks the symmetry between left and right and thus possesses a non-zero chirality. In the interactions of CPL with matter, the chirality of the electromagnetic field can couple to the chirality of matter. Various quantities have been used to describe the chirality of the electromagnetic field. A notable example is the ``zilch" introduced by D. Lipkin~\cite{Lipkin1964}: 
\begin{align}
Z^0 &= \boldsymbol{E} \cdot (\nabla\times\boldsymbol{E}) + \boldsymbol{B} \cdot (\nabla\times\boldsymbol{B})  \\
\boldsymbol{Z} &= \boldsymbol{E} \times \dot{\boldsymbol{E}} + \boldsymbol{B} \times \dot{\boldsymbol{B}}.
\end{align} Lipkin's zilch is gauge-invariant and obeys the continuity equation in free space as a consequence of Maxwell equations: $\partial_\mu Z^\mu =0$, with $Z^\mu = (Z^0, \boldsymbol{Z})$. However, interactions can transfer chirality from the electromagnetic field to matter. Chirality conservation for electromagnetic field and its role in light-matter interactions in chiral materials are the subjects of active current interest. In particular, 
Lipkin's zilch has been used to describe the interaction of CPL with chiral molecules~\cite{Tang2010,Bliokh2011,Coles2012}. CPL has also been proposed to cause a Photovoltaic Hall effect~\cite{Oka2009,Yudin2015} in graphene, which is a material with 2-dimensional  relativistic Dirac fermion quasiparticles. 

In this article, we discuss the interaction of CPL with the 3-dimensional chiral quasiparticles in recently discovered Dirac and Weyl semimetals~\cite{Taguchi2016,Ebihara2016,Chan2016,DeJuan2017,Chan2017}. We will show that the transfer of chirality from the electromagnetic field to chiral fermions can be described in a model-independent way by using the chiral anomaly~\cite{Adler1969,Bell1969}.
Because of the focus on the effect of the chiral anomaly, 
our treatment will be based on a measure of chirality that is different from Lipkin's zilch. Namely, we will use the Chern-Simons current~\cite{Chern1974}
\begin{align}
h^0 &= \boldsymbol{A}\cdot \boldsymbol{B}\\
\boldsymbol{h} &= A^0\, \boldsymbol{B} - \boldsymbol{A}\times\boldsymbol{E},
\end{align} to describe the chirality of light and its transfer to the chirality of matter. 

The Chern-Simons current is proportional to the helicity of the free electromagnetic field~\cite{Afanasiev1996}, which describes the difference between left and right circularly polarized photons. The chirality density $h^0$ is  well-known in magneto-hydrodynamics, where magnetic helicity~\cite{Woltjer1958,Moffatt1969,Arnold1998,Taylor1974} is defined as~$\int \boldsymbol{A}\cdot\boldsymbol{B}\, \mathrm{d^3}r$. Note that the helicity of the electromagnetic field is not conserved, even in free space. 

The gauge dependence of helicity is essential in describing the interactions mediated by the chiral anomaly. Indeed, the chiral anomaly results in the absence of invariance of the chiral charge $\int h^0 d^3r$ under ``large" gauge transformations that change the global topology of the gauge field. The chirality conservation law that we propose below is a consequence of the change of chirality under large gauge transformations that results from the transfer of chirality from electromagnetic field to the chiral fermion zero modes.

The chiral anomaly is known to result in the transport of charge in parallel electric and magnetic fields through the chiral magnetic effect ~\cite{Fukushima2008,Kharzeev2009} by creating a chirality imbalance. The resulting longitudinal negative magnetoresistance~\cite{son2013chiral, burkov2015negative} has been observed in Dirac semi-metals such as $\mathrm{ZrTe_5}$~\cite{Li2016} and $\mathrm{Na_3 Bi}$~\cite{Xiong2015} and Weyl semi-metals such as TaAs~\cite{Huang2015}. It has been proposed that chiral pumping in 3-dimensional Dirac materials by rotating electromagnetic field can produce a separation between left and right handed cones in momentum space, resulting in the generation of axial current and polarization of electric charge density~\cite{ebihara2016chiral}.

The interactions of light with Dirac and Weyl materials have recently attracted significant attention. A photocurrent in response to CPL, proposed for Weyl materials with tilted cones~\cite{Chan2016}, has been recently observed in TaAs~\cite{Ma2017}. A photocurrent in the presence of magnetic field has been proposed for asymmetric Weyl materials with tilted cones~\cite{Kharzeev2018a} and a quantised topological photocurrent in response to CPL has been proposed in asymmetric Weyl materials in which the left and right handed cones have different energies~\cite{DeJuan2017}. All of the above effects would be absent in Dirac materials such as $\mathrm{ZrTe_5}$ which possess both inversion and time reversal symmetries. 

In this paper, we propose a new effect that does not rely on the breaking of inversion or time reversal symmetries of the crystal and arises solely from the helicity transfer from light to the chiral fermions in an external magnetic field. The proposed effect thus provides a clean way to probe the chiral anomaly and the chiral magnetic effect. 

The effect can be briefly described as follows. If the charged chiral fermion quasiparticles are massless (the corresponding linear band has no gap), and no chirality-changing interactions are present in the Hamiltonian, the helicity of external gauge fields can be transferred to the chirality carried by the charged quasiparticles, and vice versa. The only conserved quantity is the total chirality of the fermion quasiparticles and the gauge field. 
Once the light is absorbed by the material, the helicity of the light gets fully transferred to the material. The resulting chiral imbalance between the left- and right-handed chiral fermions, as we will see, is completely fixed by the chiral anomaly. This chiral imbalance in an external magnetic field is known to induce a current due to the chiral magnetic effect ~\cite{Fukushima2008,Kharzeev2009}. Therefore, in an external magnetic field, CPL will induce an electric current.

\section{Conservation of Total Chirality}

In the interaction of CPL with an optically thick Dirac or Weyl semimetal (for a mid-infrared laser, the light penetration length for these materials is of the order of a few hundred nanometres), the helicity of the absorbed light gets fully transferred to the chirality of the fermions. 
The chiral fermions in 
Dirac and Weyl semimetals are described by the Hamiltonian which in the simplest isotropic case is given by 
\begin{equation}
\hat{H} = \pm v_\text{F} k_i \sigma_i,
\end{equation} 
where $v_\text{F}$ is the Fermi velocity and $k_i$ is the crystal momentum; the $\pm$ signs refer to the left and right handed fermions; the matrices $\sigma_i$ act over pseudospin degrees of freedom. In a Dirac material, the left and right handed cones are located at the same positions in the Brillouin zone, whereas in Weyl materials they are separated. Each Weyl cone is a monopole of the Berry curvature, and since the total Berry charge inside a Brillouin zone is zero, the left- and right-handed Weyl cones always appear in pairs.

The chirality carried by the chiral fermion quasiparticles is described by the axial current  $j_5^\mu$; the temporal component of this current is the density of chiral charge, $j_5^0 \equiv \rho_5 = \rho_\text{R} - \rho_\text{L}$. The chiral anomaly causes non-conservation of $j_5^\mu$ in the presence of an electromagnetic field $F^{\mu \nu}$, as given by~\cite{Adler1969,Bell1969,Wilczek1987,Carroll1990,Sikivie1983} 
\begin{equation}\label{chir_an}
\partial_\mu j_5^\mu = \frac{e^2}{16\pi^2} \epsilon^{\mu\nu\rho\sigma}F_{\mu\nu}F_{\rho\sigma};
\end{equation}
note that the quantity on the right hand side is odd under parity and thus vanishes for linearly polarized light.
The quantity on the right is given by the full derivative of the  Chern-Simons current~\cite{Chern1974}
\begin{equation}\label{CS}
h^\mu = -\frac{e^2}{8\pi^2}\epsilon^{\mu\nu\rho\sigma}A_\nu F_{\rho\sigma}
\end{equation}
that describes the helicity density and flux carried by the electromagnetic field. 

We deduce from~\eqref{chir_an} and~\eqref{CS} that the total chirality, which is the sum of the helicity of the electromagnetic field and the chirality of the fermions, is conserved:
\begin{equation}\label{an_cons}
\partial_\mu (j_5^\mu + h^\mu) = 0.
\end{equation}
It is this conservation law that causes transfer of chirality from light to chiral fermions. This is valid separately for each Dirac cone or pair of Weyl cones.

The helicity density and helicity flux are given (in SI units) by
\begin{equation}\label{chir_dens}
h^0 = \frac{e^2}{4\pi^2\hbar^2} \boldsymbol{A}\cdot \boldsymbol{B}
\end{equation}
and
\begin{equation}\label{chi_flux}
\boldsymbol{h} = \frac{e^2}{4\pi^2\hbar^2} (A^0\, \boldsymbol{B} - \boldsymbol{A}\times\boldsymbol{E}).
\end{equation}
In our case, the prefactor of~$e^2/(4\pi^2\hbar^2)$ appears in equation~\eqref{chir_dens} because, as it can be seen from equation~\eqref{an_cons}, it is the chirality density available for the transfer to the chiral fermions.

In a real Dirac or Weyl material, there is also chirality-flipping scattering, with a characteristic relaxation time $\tau_\mathrm{V}$, so equation~\eqref{chir_an} (in SI units) becomes
\begin{equation}\label{chir_rel}
\dot{\rho_5} + \nabla\cdot\boldsymbol{j}_5 = \frac{e^2}{2\pi^2 \hbar^2} \boldsymbol{E}\cdot\boldsymbol{B} - \frac{\rho_5}{\tau_\mathrm{V}}.
\end{equation}
The last term on the right hand side of equation~\eqref{chir_rel} does not affect the balance of chirality transfer if the frequency of light $\omega$ is large compared to ${\tau_\mathrm{V}}^{-1}$.

For an oscillating electric field  with $\boldsymbol{E}(t,\boldsymbol{r}) = \Re(\eu^{-\iu\omega t}\boldsymbol{\mathcal{E}}(\boldsymbol{r}))$, in the Coulomb gauge with $A^0 = 0$, the time-averaged helicity flux~\eqref{chi_flux} is
\begin{equation}
\langle\boldsymbol{h}\rangle = \frac{e^2}{8\pi^2\hbar^2\omega}\Re(\iu\,\boldsymbol{\mathcal{E}}\times\boldsymbol{\mathcal{E}}^*).
\end{equation}
For light traveling in the $z$~direction, this becomes
\begin{equation}
\langle h^z \rangle = \frac{e^2}{4\pi^2\hbar^2\omega}\Re(\iu\,\mathcal{E}_x \mathcal{E}_y^*);
\end{equation}
note that this quantity vanishes for linearly polarized light.

\section{Transfer of Chirality from Light to Fermions}

Let us now describe CPL using the Chern-Simons helicity. 
For CPL with $\boldsymbol{\mathcal{E}} = E_0\,(\boldsymbol{\hat{x}} \pm \iu\,\boldsymbol{\hat{y}})$,
\begin{equation}
\langle h^z \rangle = \pm \frac{e^2}{4\pi^2\hbar^2\omega} E_0^2.
\end{equation}
The ratio of this helicity flux to the energy flux $\langle S^z \rangle$ (the Poynting vector) is given by
\begin{equation}\label{per_phot}
\frac{\langle h^z \rangle}{\langle S^z \rangle} = \pm\frac{1}{\pi}\frac{e^2}{4\pi \epsilon_0 \hbar c}\frac{1}{\hbar\omega}.
\end{equation}
Hence, the helicity per photon available for transfer to each Dirac cone or pair of Weyl cones of charged fermions is~$\pm\alpha/\pi$, where~$\alpha$ is the fine structure constant; the signs refer to the two circular polarizations of light. Of course, the helicity per photon in the beam is~$\pm 1$, since photons are massless vector particles. The factor of~$\alpha/\pi$ in equation~\eqref{per_phot}, according to equation~\eqref{chir_rel}, describes the coupling of photons to the charged fermions through the chiral anomaly.  

Note that $\boldsymbol{E}\cdot\boldsymbol{B}$ is zero for CPL in vacuum, but it is non-zero if it is being attenuated in a material. This is what allows chiral charge to be generated according to~\eqref{chir_rel}.

If CPL is incident on a 3D chiral material, as the light is absorbed by the material, its helicity flux is converted into the chirality of fermions. The total chirality generated per unit area is equal to the helicity flux of light transmitted at the interface. In a real material, if $\omega \tau_\mathrm{V} \gg 1$, the chirality will saturate at a constant value proportional to $\tau_\mathrm{V}$ due to chirality relaxation, as dictated by equation~\eqref{chir_rel}. Using equation~\eqref{an_cons}, for light incident perpendicular to the interface at $z=0$,
\begin{equation}
\int_0^\infty \rho_5\,\du z= \tau_\mathrm{V} \langle h^z\rangle|_{z=0} = \pm\tau_\mathrm{V}\frac{\alpha}{\pi} \frac{I_\mathrm{in}}{\hbar \omega} \Re(a_x a_y^*),
\end{equation}
where $I_\mathrm{in}$ is the intensity of the incident light and $a_{x,y}$ are the transmission amplitudes of the two linear polarizations. The chiral charge is distributed in the material over a length scale determined by the diffusion length of fermion quasiparticles and the attenuation length of light, but the total chiral charge integrated over the depth is unaffected by this.

\begin{figure}
 \begin{center}
 \includegraphics[origin=c,width=\columnwidth]{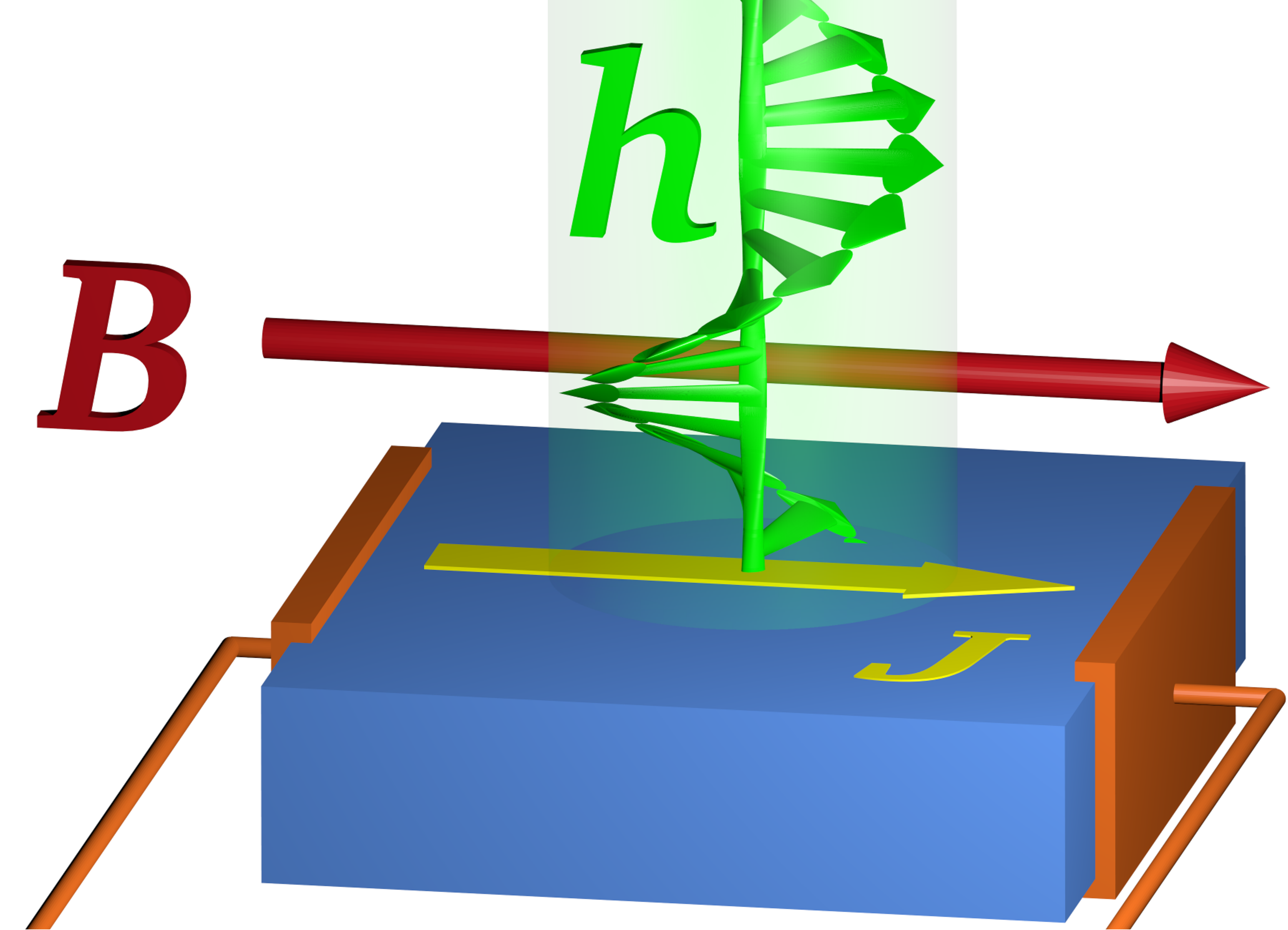}
 \caption{The helicity of circularly polarized light (CPL) is characterized by the Chern-Simons current $\boldsymbol{h}$. When incident on a Dirac or Weyl semimetal, as a consequence of the chiral anomaly, CPL induces an asymmetry between the number of left- and right-handed chiral quasiparticles. In an external magnetic field $\boldsymbol{B}$, this chiral asymmetry induces a chiral magnetic photocurrent $\boldsymbol{J}$ along the direction of the externally applied magnetic field. }
 \label{fig:expt}
 \end{center}
\end{figure}

The chiral charge density of fermions $\rho_5$ translates into a chiral chemical potential $\mu_5 \simeq \chi^{-1} \rho_5$, where $\chi=\partial \rho_5/\partial \mu_5$ is the chiral susceptibility. If the whole system is placed in a constant magnetic field $\boldsymbol{B}_\mathrm{ext}$ perpendicular to the incident light, a current 
\begin{equation}
\boldsymbol{J}_\text{CME} = \frac{e^2}{2\pi^2\hbar^2}\, \boldsymbol{B}_\mathrm{ext}\, \mu_5
\end{equation}
due to the chiral magnetic effect (CME)~\cite{Fukushima2008,Kharzeev2009} is generated along the direction of the magnetic field. The linear density $\boldsymbol{\kappa}_\text{CMP}$ of the resulting chiral magnetic photocurrent is given by the integral over the depth:
\begin{align}\label{cur_dens}
\boldsymbol{\kappa}_\text{CMP} &= \int_0^\infty \frac{e^2}{2\pi^2\hbar^2}\, \boldsymbol{B}_\text{ext}\, \mu_5\,\du z\nonumber\\& = 
\pm \frac{e^2}{2\pi^2\hbar^2}\,\boldsymbol{B}_\mathrm{ext}\,\frac{\tau_\mathrm{V}}{\chi}\,\frac{\alpha}{\pi}\, \frac{I_\mathrm{in}}{\hbar \omega}\, \Re(a_x a_y^*).
\end{align}
The formula for the CME conductivity is given by~\cite{Son2013,Xiong2015,Li2016}
\begin{align}
\left(\frac{e^2}{2\pi^2\hbar^2}\right)^2 \frac{\tau_\text{V}}{\chi} B_\text{ext}^2.
\end{align}
If the only contribution to the magnetic field dependence of the longitudinal conductivity is from the CME, then the quadratic coefficient of the longitudinal conductivity is
\begin{align}
\sigma^{(2)}_{zz} = \left(\frac{e^2}{2\pi^2\hbar^2}\right)^2 \frac{\tau_\text{V}}{\chi}
\end{align}
for each Dirac cone or pair of Weyl cones. In terms of $\sigma^{(2)}_{zz}$
\begin{equation}\label{expt_formula}
\boldsymbol{\kappa}_\text{CMP}= 
\pm \frac{2 \pi^2 \hbar^2}{e^2}\,\boldsymbol{B}_\mathrm{ext}\, \sigma^{(2)}_{zz} \,\frac{\alpha}{\pi}\, \frac{I_\mathrm{in}}{\hbar \omega}\, \Re(a_x a_y^*).
\end{equation}
This is the main result of our paper; note that this formula does not depend on the number of cones.

\section{Numerical Estimates}

The linear photocurrent density in~\eqref{expt_formula} can be integrated over the diameter of the spot of light to estimate the magnitude of the observed photocurrent. We assume a mid-infrared laser of power $10\,\mathrm{mW}$, intensity $I_\mathrm{in}=5 \times 10^{6}\, \mathrm{W/m}^2$, spot diameter $50\,\mu\mathrm{m}$,  and wavelength $10\,\mathrm{\mu m}$. We also assume that the square of the optical transmission amplitude is $\Re(a_x a_y^*)=0.1$, and use magnetic field $B_\text{ext} = 2\, \mathrm{T}$ and temperature $5\, \mathrm{K}$. We need the coefficient $\sigma^{(2)}_{zz}$ and the ratio of the resistance of the sample to the load (suppression factor) to estimate the photocurrent.

For the Weyl material TaAs used in~\cite{Zhang2016}, using the same suppression factor as the  setup in~\cite{Ma2017}, we get a current of approximately $25\,\mathrm{nA}$. For the Dirac material $\mathrm{ZrTe}_5$ used in~\cite{Li2016}, assuming the same resistance for the external load as in~\cite{Ma2017}, we get a current of approximately $50\,\mathrm{nA}$. This current scales linearly with the magnetic field and wavelength, and should be much stronger if a THz source is used.

This estimate can be compared to the photocurrent of about $40\,\mathrm{nA}$ observed recently in TaAs~\cite{Ma2017} that is said to exceed the photocurrent observed in other materials currently used for detecting the mid-infrared radiation by a factor of 10-100. 

\section{Discussion}

In contrast to the photocurrents that have been proposed~\cite{Chan2016} and observed~\cite{Ma2017} recently and the photocurrents proposed in~\cite{DeJuan2017} and~\cite{Kharzeev2018a} which utilize asymmetries of the crystals, chiral magnetic photocurrent solely depends on the imbalance between the densities of right- and left-handed chiral fermions.  The direction of the photocurrent depends only on the magnetic field and the circular polarization of light, but not on the crystal axes. The observation of the chiral magnetic photocurrent would provide a strong independent evidence for the importance of chiral anomaly in condensed matter systems.

In the effect considered in this paper, each photon flips, on average,  the chirality of $\alpha/2\pi \simeq 1/860$ fermions for \emph{each} Dirac cone, or a pair of Weyl cones. TaAs~\cite{Ma2017,Huang2015} has 12 pairs of Weyl cones, so each photon will flip the chirality of  $\simeq 1/72$ chiral fermions. The occupation number of the photons is inversely proportional to the light frequency $\omega$; because of this, the magnitude of the chiral magnetic photocurrent is proportional to $\omega^{-1}$. 

Since chiral magnetic photocurrent depends on transitions from one Weyl cone to another, instead of interband transitions, there is no lower cut-off frequency -- as a result, the photocurrent is predicted to be strong in the THz frequency range. This could potentially be used to detect circularly polarized THz radiation. In large magnetic fields and at low frequency of light, the strength of the chiral magnetic photocurrent may exceed the currently observed chiral photocurrents. This opens possibilities for applications in photonics and optoelectronics, especially in the THz frequency range.

\section{Acknowledgement}
We thank Qiang Li, Xu Du, Mengkun Liu, Yuta Kikuchi, and Ren\'e Meyer for useful and stimulating discussions. We also thank the referees for their suggestions and constructive criticism. This work was supported in part by the U.S. Department of Energy under Awards DE-SC-0017662, DE-FG-88ER40388, and DE-AC02-98CH10886.

\bibliography{MagneticPhotocurrent_bib}

\begin{thebibliography}{34}%
\makeatletter
\providecommand \@ifxundefined [1]{%
 \@ifx{#1\undefined}
}%
\providecommand \@ifnum [1]{%
 \ifnum #1\expandafter \@firstoftwo
 \else \expandafter \@secondoftwo
 \fi
}%
\providecommand \@ifx [1]{%
 \ifx #1\expandafter \@firstoftwo
 \else \expandafter \@secondoftwo
 \fi
}%
\providecommand \natexlab [1]{#1}%
\providecommand \enquote  [1]{``#1''}%
\providecommand \bibnamefont  [1]{#1}%
\providecommand \bibfnamefont [1]{#1}%
\providecommand \citenamefont [1]{#1}%
\providecommand \href@noop [0]{\@secondoftwo}%
\providecommand \href [0]{\begingroup \@sanitize@url \@href}%
\providecommand \@href[1]{\@@startlink{#1}\@@href}%
\providecommand \@@href[1]{\endgroup#1\@@endlink}%
\providecommand \@sanitize@url [0]{\catcode `\\12\catcode `\$12\catcode
  `\&12\catcode `\#12\catcode `\^12\catcode `\_12\catcode `\%12\relax}%
\providecommand \@@startlink[1]{}%
\providecommand \@@endlink[0]{}%
\providecommand \url  [0]{\begingroup\@sanitize@url \@url }%
\providecommand \@url [1]{\endgroup\@href {#1}{\urlprefix }}%
\providecommand \urlprefix  [0]{URL }%
\providecommand \Eprint [0]{\href }%
\providecommand \doibase [0]{http://dx.doi.org/}%
\providecommand \selectlanguage [0]{\@gobble}%
\providecommand \bibinfo  [0]{\@secondoftwo}%
\providecommand \bibfield  [0]{\@secondoftwo}%
\providecommand \translation [1]{[#1]}%
\providecommand \BibitemOpen [0]{}%
\providecommand \bibitemStop [0]{}%
\providecommand \bibitemNoStop [0]{.\EOS\space}%
\providecommand \EOS [0]{\spacefactor3000\relax}%
\providecommand \BibitemShut  [1]{\csname bibitem#1\endcsname}%
\let\auto@bib@innerbib\@empty
\bibitem [{\citenamefont {Lipkin}(1964)}]{Lipkin1964}%
  \BibitemOpen
  \bibfield  {author} {\bibinfo {author} {\bibfnamefont {D.~M.}\ \bibnamefont
  {Lipkin}},\ }\href {\doibase 10.1063/1.1704165} {\bibfield  {journal}
  {\bibinfo  {journal} {Journal of Mathematical Physics}\ }\textbf {\bibinfo
  {volume} {5}},\ \bibinfo {pages} {696} (\bibinfo {year} {1964})}\BibitemShut
  {NoStop}%
\bibitem [{\citenamefont {Tang}\ and\ \citenamefont {Cohen}(2010)}]{Tang2010}%
  \BibitemOpen
  \bibfield  {author} {\bibinfo {author} {\bibfnamefont {Y.}~\bibnamefont
  {Tang}}\ and\ \bibinfo {author} {\bibfnamefont {A.~E.}\ \bibnamefont
  {Cohen}},\ }\href {\doibase 10.1103/PhysRevLett.104.163901} {\bibfield
  {journal} {\bibinfo  {journal} {Physical Review Letters}\ }\textbf {\bibinfo
  {volume} {104}},\ \bibinfo {pages} {163901} (\bibinfo {year}
  {2010})}\BibitemShut {NoStop}%
\bibitem [{\citenamefont {Bliokh}\ and\ \citenamefont
  {Nori}(2011)}]{Bliokh2011}%
  \BibitemOpen
  \bibfield  {author} {\bibinfo {author} {\bibfnamefont {K.~Y.}\ \bibnamefont
  {Bliokh}}\ and\ \bibinfo {author} {\bibfnamefont {F.}~\bibnamefont {Nori}},\
  }\href {\doibase 10.1103/PhysRevA.83.021803} {\bibfield  {journal} {\bibinfo
  {journal} {Physical Review A}\ }\textbf {\bibinfo {volume} {83}},\ \bibinfo
  {pages} {021803} (\bibinfo {year} {2011})},\ \Eprint
  {http://arxiv.org/abs/1012.4176} {arXiv:1012.4176 [physics.optics]}
  \BibitemShut {NoStop}%
\bibitem [{\citenamefont {Coles}\ and\ \citenamefont
  {Andrews}(2012)}]{Coles2012}%
  \BibitemOpen
  \bibfield  {author} {\bibinfo {author} {\bibfnamefont {M.~M.}\ \bibnamefont
  {Coles}}\ and\ \bibinfo {author} {\bibfnamefont {D.~L.}\ \bibnamefont
  {Andrews}},\ }\href {\doibase 10.1103/PhysRevA.85.063810} {\bibfield
  {journal} {\bibinfo  {journal} {Physical Review A}\ }\textbf {\bibinfo
  {volume} {85}},\ \bibinfo {pages} {063810} (\bibinfo {year} {2012})},\
  \Eprint {http://arxiv.org/abs/1203.1755} {arXiv:1203.1755 [physics.optics]}
  \BibitemShut {NoStop}%
\bibitem [{\citenamefont {Oka}\ and\ \citenamefont {Aoki}(2009)}]{Oka2009}%
  \BibitemOpen
  \bibfield  {author} {\bibinfo {author} {\bibfnamefont {T.}~\bibnamefont
  {Oka}}\ and\ \bibinfo {author} {\bibfnamefont {H.}~\bibnamefont {Aoki}},\
  }\href {\doibase 10.1103/PhysRevB.79.081406} {\bibfield  {journal} {\bibinfo
  {journal} {Physical Review B}\ }\textbf {\bibinfo {volume} {79}},\ \bibinfo
  {pages} {081406} (\bibinfo {year} {2009})},\ \Eprint
  {http://arxiv.org/abs/0807.4767} {arXiv:0807.4767 [cond-mat.mes-hall]}
  \BibitemShut {NoStop}%
\bibitem [{\citenamefont {Yudin}\ \emph {et~al.}(2015)\citenamefont {Yudin},
  \citenamefont {Eriksson},\ and\ \citenamefont {Katsnelson}}]{Yudin2015}%
  \BibitemOpen
  \bibfield  {author} {\bibinfo {author} {\bibfnamefont {D.}~\bibnamefont
  {Yudin}}, \bibinfo {author} {\bibfnamefont {O.}~\bibnamefont {Eriksson}}, \
  and\ \bibinfo {author} {\bibfnamefont {M.~I.}\ \bibnamefont {Katsnelson}},\
  }\href {\doibase 10.1103/PhysRevB.91.075419} {\bibfield  {journal} {\bibinfo
  {journal} {Physical Review B}\ }\textbf {\bibinfo {volume} {91}},\ \bibinfo
  {pages} {075419} (\bibinfo {year} {2015})},\ \Eprint
  {http://arxiv.org/abs/1406.3980} {arXiv:1406.3980 [cond-mat.mes-hall]}
  \BibitemShut {NoStop}%
\bibitem [{\citenamefont {Taguchi}\ \emph {et~al.}(2016)\citenamefont
  {Taguchi}, \citenamefont {Imaeda}, \citenamefont {Sato},\ and\ \citenamefont
  {Tanaka}}]{Taguchi2016}%
  \BibitemOpen
  \bibfield  {author} {\bibinfo {author} {\bibfnamefont {K.}~\bibnamefont
  {Taguchi}}, \bibinfo {author} {\bibfnamefont {T.}~\bibnamefont {Imaeda}},
  \bibinfo {author} {\bibfnamefont {M.}~\bibnamefont {Sato}}, \ and\ \bibinfo
  {author} {\bibfnamefont {Y.}~\bibnamefont {Tanaka}},\ }\href {\doibase
  10.1103/PhysRevB.93.201202} {\bibfield  {journal} {\bibinfo  {journal}
  {Physical Review B}\ }\textbf {\bibinfo {volume} {93}},\ \bibinfo {pages}
  {201202} (\bibinfo {year} {2016})},\ \Eprint
  {http://arxiv.org/abs/1601.00379} {arXiv:1601.00379 [cond-mat.mes-hall]}
  \BibitemShut {NoStop}%
\bibitem [{\citenamefont {Ebihara}\ \emph
  {et~al.}(2016{\natexlab{a}})\citenamefont {Ebihara}, \citenamefont
  {Fukushima},\ and\ \citenamefont {Oka}}]{Ebihara2016}%
  \BibitemOpen
  \bibfield  {author} {\bibinfo {author} {\bibfnamefont {S.}~\bibnamefont
  {Ebihara}}, \bibinfo {author} {\bibfnamefont {K.}~\bibnamefont {Fukushima}},
  \ and\ \bibinfo {author} {\bibfnamefont {T.}~\bibnamefont {Oka}},\ }\href
  {\doibase 10.1103/PhysRevB.93.155107} {\bibfield  {journal} {\bibinfo
  {journal} {Physical Review B}\ }\textbf {\bibinfo {volume} {93}},\ \bibinfo
  {pages} {155107} (\bibinfo {year} {2016}{\natexlab{a}})},\ \Eprint
  {http://arxiv.org/abs/1509.03673} {arXiv:1509.03673 [cond-mat.str-el]}
  \BibitemShut {NoStop}%
\bibitem [{\citenamefont {Chan}\ \emph {et~al.}(2016)\citenamefont {Chan},
  \citenamefont {Lee}, \citenamefont {Burch}, \citenamefont {Han},\ and\
  \citenamefont {Ran}}]{Chan2016}%
  \BibitemOpen
  \bibfield  {author} {\bibinfo {author} {\bibfnamefont {C.-K.}\ \bibnamefont
  {Chan}}, \bibinfo {author} {\bibfnamefont {P.~A.}\ \bibnamefont {Lee}},
  \bibinfo {author} {\bibfnamefont {K.~S.}\ \bibnamefont {Burch}}, \bibinfo
  {author} {\bibfnamefont {J.~H.}\ \bibnamefont {Han}}, \ and\ \bibinfo
  {author} {\bibfnamefont {Y.}~\bibnamefont {Ran}},\ }\href {\doibase
  10.1103/PhysRevLett.116.026805} {\bibfield  {journal} {\bibinfo  {journal}
  {Physical Review Letters}\ }\textbf {\bibinfo {volume} {116}},\ \bibinfo
  {pages} {026805} (\bibinfo {year} {2016})},\ \Eprint
  {http://arxiv.org/abs/1509.05400} {arXiv:1509.05400 [cond-mat.mes-hall]}
  \BibitemShut {NoStop}%
\bibitem [{\citenamefont {de~Juan}\ \emph {et~al.}(2017)\citenamefont
  {de~Juan}, \citenamefont {Grushin}, \citenamefont {Morimoto},\ and\
  \citenamefont {Moore}}]{DeJuan2017}%
  \BibitemOpen
  \bibfield  {author} {\bibinfo {author} {\bibfnamefont {F.}~\bibnamefont
  {de~Juan}}, \bibinfo {author} {\bibfnamefont {A.~G.}\ \bibnamefont
  {Grushin}}, \bibinfo {author} {\bibfnamefont {T.}~\bibnamefont {Morimoto}}, \
  and\ \bibinfo {author} {\bibfnamefont {J.~E.}\ \bibnamefont {Moore}},\ }\href
  {\doibase 10.1038/ncomms15995} {\bibfield  {journal} {\bibinfo  {journal}
  {Nature Communications}\ }\textbf {\bibinfo {volume} {8}},\ \bibinfo {pages}
  {15995} (\bibinfo {year} {2017})},\ \Eprint {http://arxiv.org/abs/1611.05887}
  {arXiv:1611.05887 [cond-mat.str-el]} \BibitemShut {NoStop}%
\bibitem [{\citenamefont {Chan}\ \emph {et~al.}(2017)\citenamefont {Chan},
  \citenamefont {Lindner}, \citenamefont {Refael},\ and\ \citenamefont
  {Lee}}]{Chan2017}%
  \BibitemOpen
  \bibfield  {author} {\bibinfo {author} {\bibfnamefont {C.-K.}\ \bibnamefont
  {Chan}}, \bibinfo {author} {\bibfnamefont {N.~H.}\ \bibnamefont {Lindner}},
  \bibinfo {author} {\bibfnamefont {G.}~\bibnamefont {Refael}}, \ and\ \bibinfo
  {author} {\bibfnamefont {P.~A.}\ \bibnamefont {Lee}},\ }\href {\doibase
  10.1103/PhysRevB.95.041104} {\bibfield  {journal} {\bibinfo  {journal}
  {Physical Review B}\ }\textbf {\bibinfo {volume} {95}},\ \bibinfo {pages}
  {041104} (\bibinfo {year} {2017})},\ \Eprint
  {http://arxiv.org/abs/1607.07839} {arXiv:1607.07839 [cond-mat.mes-hall]}
  \BibitemShut {NoStop}%
\bibitem [{\citenamefont {Adler}(1969)}]{Adler1969}%
  \BibitemOpen
  \bibfield  {author} {\bibinfo {author} {\bibfnamefont {S.~L.}\ \bibnamefont
  {Adler}},\ }\href {\doibase 10.1103/PhysRev.177.2426} {\bibfield  {journal}
  {\bibinfo  {journal} {Physical Review}\ }\textbf {\bibinfo {volume} {177}},\
  \bibinfo {pages} {2426} (\bibinfo {year} {1969})}\BibitemShut {NoStop}%
\bibitem [{\citenamefont {Bell}\ and\ \citenamefont {Jackiw}(1969)}]{Bell1969}%
  \BibitemOpen
  \bibfield  {author} {\bibinfo {author} {\bibfnamefont {J.~S.}\ \bibnamefont
  {Bell}}\ and\ \bibinfo {author} {\bibfnamefont {R.}~\bibnamefont {Jackiw}},\
  }\href {\doibase 10.1007/BF02823296} {\bibfield  {journal} {\bibinfo
  {journal} {Il Nuovo Cimento A}\ }\textbf {\bibinfo {volume} {60}},\ \bibinfo
  {pages} {47} (\bibinfo {year} {1969})}\BibitemShut {NoStop}%
\bibitem [{\citenamefont {Chern}\ and\ \citenamefont
  {Simons}(1974)}]{Chern1974}%
  \BibitemOpen
  \bibfield  {author} {\bibinfo {author} {\bibfnamefont {S.-S.}\ \bibnamefont
  {Chern}}\ and\ \bibinfo {author} {\bibfnamefont {J.}~\bibnamefont {Simons}},\
  }\href {\doibase 10.2307/1971013} {\bibfield  {journal} {\bibinfo  {journal}
  {The Annals of Mathematics}\ }\textbf {\bibinfo {volume} {99}},\ \bibinfo
  {pages} {48} (\bibinfo {year} {1974})}\BibitemShut {NoStop}%
\bibitem [{\citenamefont {Afanasiev}\ and\ \citenamefont
  {Stepanovsky}(1996)}]{Afanasiev1996}%
  \BibitemOpen
  \bibfield  {author} {\bibinfo {author} {\bibfnamefont {G.~N.}\ \bibnamefont
  {Afanasiev}}\ and\ \bibinfo {author} {\bibfnamefont {Y.~P.}\ \bibnamefont
  {Stepanovsky}},\ }\href {\doibase 10.1007/BF02731014} {\bibfield  {journal}
  {\bibinfo  {journal} {Il Nuovo Cimento A}\ }\textbf {\bibinfo {volume}
  {109}},\ \bibinfo {pages} {271} (\bibinfo {year} {1996})}\BibitemShut
  {NoStop}%
\bibitem [{\citenamefont {Woltjer}(1958)}]{Woltjer1958}%
  \BibitemOpen
  \bibfield  {author} {\bibinfo {author} {\bibfnamefont {L.}~\bibnamefont
  {Woltjer}},\ }\href
  {http://www.pubmedcentral.nih.gov/articlerender.fcgi?artid=PMC528606}
  {\bibfield  {journal} {\bibinfo  {journal} {Proceedings of the National
  Academy of Sciences of the United States of America}\ }\textbf {\bibinfo
  {volume} {44}},\ \bibinfo {pages} {489} (\bibinfo {year} {1958})}\BibitemShut
  {NoStop}%
\bibitem [{\citenamefont {Moffatt}(1969)}]{Moffatt1969}%
  \BibitemOpen
  \bibfield  {author} {\bibinfo {author} {\bibfnamefont {H.~K.}\ \bibnamefont
  {Moffatt}},\ }\href {\doibase 10.1017/S0022112069000991} {\bibfield
  {journal} {\bibinfo  {journal} {Journal of Fluid Mechanics}\ }\textbf
  {\bibinfo {volume} {35}},\ \bibinfo {pages} {117} (\bibinfo {year}
  {1969})}\BibitemShut {NoStop}%
\bibitem [{\citenamefont {Arnold}\ and\ \citenamefont
  {Khesin}(1998)}]{Arnold1998}%
  \BibitemOpen
  \bibfield  {author} {\bibinfo {author} {\bibfnamefont {V.~I.}\ \bibnamefont
  {Arnold}}\ and\ \bibinfo {author} {\bibfnamefont {B.~A.}\ \bibnamefont
  {Khesin}},\ }\href {\doibase 10.1007/b97593} {\emph {\bibinfo {title}
  {{Topological Methods in Hydrodynamics}}}},\ \bibinfo {series} {Applied
  Mathematical Sciences}, Vol.\ \bibinfo {volume} {125}\ (\bibinfo  {publisher}
  {Springer New York},\ \bibinfo {address} {New York, NY},\ \bibinfo {year}
  {1998})\BibitemShut {NoStop}%
\bibitem [{\citenamefont {Taylor}(1974)}]{Taylor1974}%
  \BibitemOpen
  \bibfield  {author} {\bibinfo {author} {\bibfnamefont {J.~B.}\ \bibnamefont
  {Taylor}},\ }\href {\doibase 10.1103/PhysRevLett.33.1139} {\bibfield
  {journal} {\bibinfo  {journal} {Physical Review Letters}\ }\textbf {\bibinfo
  {volume} {33}},\ \bibinfo {pages} {1139} (\bibinfo {year}
  {1974})}\BibitemShut {NoStop}%
\bibitem [{\citenamefont {Fukushima}\ \emph {et~al.}(2008)\citenamefont
  {Fukushima}, \citenamefont {Kharzeev},\ and\ \citenamefont
  {Warringa}}]{Fukushima2008}%
  \BibitemOpen
  \bibfield  {author} {\bibinfo {author} {\bibfnamefont {K.}~\bibnamefont
  {Fukushima}}, \bibinfo {author} {\bibfnamefont {D.~E.}\ \bibnamefont
  {Kharzeev}}, \ and\ \bibinfo {author} {\bibfnamefont {H.~J.}\ \bibnamefont
  {Warringa}},\ }\href {\doibase 10.1103/PhysRevD.78.074033} {\bibfield
  {journal} {\bibinfo  {journal} {Physical Review D}\ }\textbf {\bibinfo
  {volume} {78}},\ \bibinfo {pages} {074033} (\bibinfo {year} {2008})},\
  \Eprint {http://arxiv.org/abs/1002.2495} {arXiv:1002.2495 [hep-ph]}
  \BibitemShut {NoStop}%
\bibitem [{\citenamefont {Kharzeev}\ and\ \citenamefont
  {Warringa}(2009)}]{Kharzeev2009}%
  \BibitemOpen
  \bibfield  {author} {\bibinfo {author} {\bibfnamefont {D.~E.}\ \bibnamefont
  {Kharzeev}}\ and\ \bibinfo {author} {\bibfnamefont {H.~J.}\ \bibnamefont
  {Warringa}},\ }\href {\doibase 10.1103/PhysRevD.80.034028} {\bibfield
  {journal} {\bibinfo  {journal} {Physical Review D}\ }\textbf {\bibinfo
  {volume} {80}},\ \bibinfo {pages} {034028} (\bibinfo {year} {2009})},\
  \Eprint {http://arxiv.org/abs/0907.5007} {arXiv:0907.5007 [hep-ph]}
  \BibitemShut {NoStop}%
\bibitem [{\citenamefont {Son}\ and\ \citenamefont
  {Spivak}(2013{\natexlab{a}})}]{son2013chiral}%
  \BibitemOpen
  \bibfield  {author} {\bibinfo {author} {\bibfnamefont {D.~T.}\ \bibnamefont
  {Son}}\ and\ \bibinfo {author} {\bibfnamefont {B.~Z.}\ \bibnamefont
  {Spivak}},\ }\href {\doibase 10.1103/PhysRevB.88.104412} {\bibfield
  {journal} {\bibinfo  {journal} {Phys. Rev.}\ }\textbf {\bibinfo {volume}
  {B88}},\ \bibinfo {pages} {104412} (\bibinfo {year} {2013}{\natexlab{a}})},\
  \Eprint {http://arxiv.org/abs/1206.1627} {arXiv:1206.1627
  [cond-mat.mes-hall]} \BibitemShut {NoStop}%
\bibitem [{\citenamefont {Burkov}(2015)}]{burkov2015negative}%
  \BibitemOpen
  \bibfield  {author} {\bibinfo {author} {\bibfnamefont {A.~A.}\ \bibnamefont
  {Burkov}},\ }\href {\doibase 10.1103/PhysRevB.91.245157} {\bibfield
  {journal} {\bibinfo  {journal} {Phys. Rev. B}\ }\textbf {\bibinfo {volume}
  {91}},\ \bibinfo {pages} {245157} (\bibinfo {year} {2015})},\ \Eprint
  {http://arxiv.org/abs/1505.01849} {arXiv:1505.01849 [cond-mat.mes-hall]}
  \BibitemShut {NoStop}%
\bibitem [{\citenamefont {Li}\ \emph {et~al.}(2016)\citenamefont {Li},
  \citenamefont {Kharzeev}, \citenamefont {Zhang}, \citenamefont {Huang},
  \citenamefont {Pletikosi{\'{c}}}, \citenamefont {Fedorov}, \citenamefont
  {Zhong}, \citenamefont {Schneeloch}, \citenamefont {Gu},\ and\ \citenamefont
  {Valla}}]{Li2016}%
  \BibitemOpen
  \bibfield  {author} {\bibinfo {author} {\bibfnamefont {Q.}~\bibnamefont
  {Li}}, \bibinfo {author} {\bibfnamefont {D.~E.}\ \bibnamefont {Kharzeev}},
  \bibinfo {author} {\bibfnamefont {C.}~\bibnamefont {Zhang}}, \bibinfo
  {author} {\bibfnamefont {Y.}~\bibnamefont {Huang}}, \bibinfo {author}
  {\bibfnamefont {I.}~\bibnamefont {Pletikosi{\'{c}}}}, \bibinfo {author}
  {\bibfnamefont {A.~V.}\ \bibnamefont {Fedorov}}, \bibinfo {author}
  {\bibfnamefont {R.~D.}\ \bibnamefont {Zhong}}, \bibinfo {author}
  {\bibfnamefont {J.~A.}\ \bibnamefont {Schneeloch}}, \bibinfo {author}
  {\bibfnamefont {G.~D.}\ \bibnamefont {Gu}}, \ and\ \bibinfo {author}
  {\bibfnamefont {T.}~\bibnamefont {Valla}},\ }\href {\doibase
  10.1038/nphys3648} {\bibfield  {journal} {\bibinfo  {journal} {Nature
  Physics}\ }\textbf {\bibinfo {volume} {12}},\ \bibinfo {pages} {550}
  (\bibinfo {year} {2016})},\ \Eprint {http://arxiv.org/abs/1412.6543}
  {arXiv:1412.6543 [cond-mat.str-el]} \BibitemShut {NoStop}%
\bibitem [{\citenamefont {Xiong}\ \emph {et~al.}(2015)\citenamefont {Xiong},
  \citenamefont {Kushwaha}, \citenamefont {Liang}, \citenamefont {Krizan},
  \citenamefont {Hirschberger}, \citenamefont {Wang}, \citenamefont {Cava},\
  and\ \citenamefont {Ong}}]{Xiong2015}%
  \BibitemOpen
  \bibfield  {author} {\bibinfo {author} {\bibfnamefont {J.}~\bibnamefont
  {Xiong}}, \bibinfo {author} {\bibfnamefont {S.~K.}\ \bibnamefont {Kushwaha}},
  \bibinfo {author} {\bibfnamefont {T.}~\bibnamefont {Liang}}, \bibinfo
  {author} {\bibfnamefont {J.~W.}\ \bibnamefont {Krizan}}, \bibinfo {author}
  {\bibfnamefont {M.}~\bibnamefont {Hirschberger}}, \bibinfo {author}
  {\bibfnamefont {W.}~\bibnamefont {Wang}}, \bibinfo {author} {\bibfnamefont
  {R.~J.}\ \bibnamefont {Cava}}, \ and\ \bibinfo {author} {\bibfnamefont
  {N.~P.}\ \bibnamefont {Ong}},\ }\href {\doibase 10.1126/science.aac6089}
  {\bibfield  {journal} {\bibinfo  {journal} {Science}\ }\textbf {\bibinfo
  {volume} {350}},\ \bibinfo {pages} {413} (\bibinfo {year}
  {2015})}\BibitemShut {NoStop}%
\bibitem [{\citenamefont {Huang}\ \emph {et~al.}(2015)\citenamefont {Huang},
  \citenamefont {Zhao}, \citenamefont {Long}, \citenamefont {Wang},
  \citenamefont {Chen}, \citenamefont {Yang}, \citenamefont {Liang},
  \citenamefont {Xue}, \citenamefont {Weng}, \citenamefont {Fang},
  \citenamefont {Dai},\ and\ \citenamefont {Chen}}]{Huang2015}%
  \BibitemOpen
  \bibfield  {author} {\bibinfo {author} {\bibfnamefont {X.}~\bibnamefont
  {Huang}}, \bibinfo {author} {\bibfnamefont {L.}~\bibnamefont {Zhao}},
  \bibinfo {author} {\bibfnamefont {Y.}~\bibnamefont {Long}}, \bibinfo {author}
  {\bibfnamefont {P.}~\bibnamefont {Wang}}, \bibinfo {author} {\bibfnamefont
  {D.}~\bibnamefont {Chen}}, \bibinfo {author} {\bibfnamefont {Z.}~\bibnamefont
  {Yang}}, \bibinfo {author} {\bibfnamefont {H.}~\bibnamefont {Liang}},
  \bibinfo {author} {\bibfnamefont {M.}~\bibnamefont {Xue}}, \bibinfo {author}
  {\bibfnamefont {H.}~\bibnamefont {Weng}}, \bibinfo {author} {\bibfnamefont
  {Z.}~\bibnamefont {Fang}}, \bibinfo {author} {\bibfnamefont {X.}~\bibnamefont
  {Dai}}, \ and\ \bibinfo {author} {\bibfnamefont {G.}~\bibnamefont {Chen}},\
  }\href {\doibase 10.1103/PhysRevX.5.031023} {\bibfield  {journal} {\bibinfo
  {journal} {Physical Review X}\ }\textbf {\bibinfo {volume} {5}},\ \bibinfo
  {pages} {031023} (\bibinfo {year} {2015})},\ \Eprint
  {http://arxiv.org/abs/1503.01304} {arXiv:1503.01304 [cond-mat.mtrl-sci]}
  \BibitemShut {NoStop}%
\bibitem [{\citenamefont {Ebihara}\ \emph
  {et~al.}(2016{\natexlab{b}})\citenamefont {Ebihara}, \citenamefont
  {Fukushima},\ and\ \citenamefont {Oka}}]{ebihara2016chiral}%
  \BibitemOpen
  \bibfield  {author} {\bibinfo {author} {\bibfnamefont {S.}~\bibnamefont
  {Ebihara}}, \bibinfo {author} {\bibfnamefont {K.}~\bibnamefont {Fukushima}},
  \ and\ \bibinfo {author} {\bibfnamefont {T.}~\bibnamefont {Oka}},\ }\href
  {\doibase 10.1103/PhysRevB.93.155107} {\bibfield  {journal} {\bibinfo
  {journal} {Phys. Rev.}\ }\textbf {\bibinfo {volume} {B93}},\ \bibinfo {pages}
  {155107} (\bibinfo {year} {2016}{\natexlab{b}})},\ \Eprint
  {http://arxiv.org/abs/1509.03673} {arXiv:1509.03673 [cond-mat.str-el]}
  \BibitemShut {NoStop}%
\bibitem [{\citenamefont {Ma}\ \emph {et~al.}(2017)\citenamefont {Ma},
  \citenamefont {Xu}, \citenamefont {Chan}, \citenamefont {Zhang},
  \citenamefont {Chang}, \citenamefont {Lin}, \citenamefont {Xie},
  \citenamefont {Palacios}, \citenamefont {Lin}, \citenamefont {Jia},
  \citenamefont {Lee}, \citenamefont {Jarillo-Herrero},\ and\ \citenamefont
  {Gedik}}]{Ma2017}%
  \BibitemOpen
  \bibfield  {author} {\bibinfo {author} {\bibfnamefont {Q.}~\bibnamefont
  {Ma}}, \bibinfo {author} {\bibfnamefont {S.-Y.}\ \bibnamefont {Xu}}, \bibinfo
  {author} {\bibfnamefont {C.-K.}\ \bibnamefont {Chan}}, \bibinfo {author}
  {\bibfnamefont {C.-L.}\ \bibnamefont {Zhang}}, \bibinfo {author}
  {\bibfnamefont {G.}~\bibnamefont {Chang}}, \bibinfo {author} {\bibfnamefont
  {Y.}~\bibnamefont {Lin}}, \bibinfo {author} {\bibfnamefont {W.}~\bibnamefont
  {Xie}}, \bibinfo {author} {\bibfnamefont {T.}~\bibnamefont {Palacios}},
  \bibinfo {author} {\bibfnamefont {H.}~\bibnamefont {Lin}}, \bibinfo {author}
  {\bibfnamefont {S.}~\bibnamefont {Jia}}, \bibinfo {author} {\bibfnamefont
  {P.~A.}\ \bibnamefont {Lee}}, \bibinfo {author} {\bibfnamefont
  {P.}~\bibnamefont {Jarillo-Herrero}}, \ and\ \bibinfo {author} {\bibfnamefont
  {N.}~\bibnamefont {Gedik}},\ }\href {\doibase 10.1038/nphys4146} {\bibfield
  {journal} {\bibinfo  {journal} {Nature Physics}\ }\textbf {\bibinfo {volume}
  {13}},\ \bibinfo {pages} {842} (\bibinfo {year} {2017})},\ \Eprint
  {http://arxiv.org/abs/1705.00590} {arXiv:1705.00590 [cond-mat.mtrl-sci]}
  \BibitemShut {NoStop}%
\bibitem [{\citenamefont {Kharzeev}\ \emph {et~al.}(2018)\citenamefont
  {Kharzeev}, \citenamefont {Kikuchi}, \citenamefont {Meyer},\ and\
  \citenamefont {Tanizaki}}]{Kharzeev2018a}%
  \BibitemOpen
  \bibfield  {author} {\bibinfo {author} {\bibfnamefont {D.~E.}\ \bibnamefont
  {Kharzeev}}, \bibinfo {author} {\bibfnamefont {Y.}~\bibnamefont {Kikuchi}},
  \bibinfo {author} {\bibfnamefont {R.}~\bibnamefont {Meyer}}, \ and\ \bibinfo
  {author} {\bibfnamefont {Y.}~\bibnamefont {Tanizaki}},\ }\href {\doibase
  10.1103/PhysRevB.98.014305} {\bibfield  {journal} {\bibinfo  {journal}
  {Physical Review B}\ }\textbf {\bibinfo {volume} {98}},\ \bibinfo {pages}
  {014305} (\bibinfo {year} {2018})},\ \Eprint
  {http://arxiv.org/abs/1801.10283} {arXiv:1801.10283 [cond-mat.mes-hall]}
  \BibitemShut {NoStop}%
\bibitem [{\citenamefont {Wilczek}(1987)}]{Wilczek1987}%
  \BibitemOpen
  \bibfield  {author} {\bibinfo {author} {\bibfnamefont {F.}~\bibnamefont
  {Wilczek}},\ }\href {\doibase 10.1103/PhysRevLett.58.1799} {\bibfield
  {journal} {\bibinfo  {journal} {Physical Review Letters}\ }\textbf {\bibinfo
  {volume} {58}},\ \bibinfo {pages} {1799} (\bibinfo {year}
  {1987})}\BibitemShut {NoStop}%
\bibitem [{\citenamefont {Carroll}\ \emph {et~al.}(1990)\citenamefont
  {Carroll}, \citenamefont {Field},\ and\ \citenamefont
  {Jackiw}}]{Carroll1990}%
  \BibitemOpen
  \bibfield  {author} {\bibinfo {author} {\bibfnamefont {S.~M.}\ \bibnamefont
  {Carroll}}, \bibinfo {author} {\bibfnamefont {G.~B.}\ \bibnamefont {Field}},
  \ and\ \bibinfo {author} {\bibfnamefont {R.}~\bibnamefont {Jackiw}},\ }\href
  {\doibase 10.1103/PhysRevD.41.1231} {\bibfield  {journal} {\bibinfo
  {journal} {Physical Review D}\ }\textbf {\bibinfo {volume} {41}},\ \bibinfo
  {pages} {1231} (\bibinfo {year} {1990})}\BibitemShut {NoStop}%
\bibitem [{\citenamefont {Sikivie}(1983)}]{Sikivie1983}%
  \BibitemOpen
  \bibfield  {author} {\bibinfo {author} {\bibfnamefont {P.}~\bibnamefont
  {Sikivie}},\ }\href {\doibase 10.1103/PhysRevLett.51.1415} {\bibfield
  {journal} {\bibinfo  {journal} {Physical Review Letters}\ }\textbf {\bibinfo
  {volume} {51}},\ \bibinfo {pages} {1415} (\bibinfo {year}
  {1983})}\BibitemShut {NoStop}%
\bibitem [{\citenamefont {Son}\ and\ \citenamefont
  {Spivak}(2013{\natexlab{b}})}]{Son2013}%
  \BibitemOpen
  \bibfield  {author} {\bibinfo {author} {\bibfnamefont {D.~T.}\ \bibnamefont
  {Son}}\ and\ \bibinfo {author} {\bibfnamefont {B.~Z.}\ \bibnamefont
  {Spivak}},\ }\href {\doibase 10.1103/PhysRevB.88.104412} {\bibfield
  {journal} {\bibinfo  {journal} {Physical Review B}\ }\textbf {\bibinfo
  {volume} {88}},\ \bibinfo {pages} {104412} (\bibinfo {year}
  {2013}{\natexlab{b}})},\ \Eprint {http://arxiv.org/abs/1206.1627}
  {arXiv:1206.1627 [cond-mat.mes-hall]} \BibitemShut {NoStop}%
\bibitem [{\citenamefont {Zhang}\ \emph {et~al.}(2016)\citenamefont {Zhang},
  \citenamefont {Xu}, \citenamefont {Belopolski}, \citenamefont {Yuan},
  \citenamefont {Lin}, \citenamefont {Tong}, \citenamefont {Bian},
  \citenamefont {Alidoust}, \citenamefont {Lee}, \citenamefont {Huang},
  \citenamefont {Chang}, \citenamefont {Chang}, \citenamefont {Hsu},
  \citenamefont {Jeng}, \citenamefont {Neupane}, \citenamefont {Sanchez},
  \citenamefont {Zheng}, \citenamefont {Wang}, \citenamefont {Lin},
  \citenamefont {Zhang}, \citenamefont {Lu}, \citenamefont {Shen},
  \citenamefont {Neupert}, \citenamefont {{Zahid Hasan}},\ and\ \citenamefont
  {Jia}}]{Zhang2016}%
  \BibitemOpen
  \bibfield  {author} {\bibinfo {author} {\bibfnamefont {C.-L.}\ \bibnamefont
  {Zhang}}, \bibinfo {author} {\bibfnamefont {S.-Y.}\ \bibnamefont {Xu}},
  \bibinfo {author} {\bibfnamefont {I.}~\bibnamefont {Belopolski}}, \bibinfo
  {author} {\bibfnamefont {Z.}~\bibnamefont {Yuan}}, \bibinfo {author}
  {\bibfnamefont {Z.}~\bibnamefont {Lin}}, \bibinfo {author} {\bibfnamefont
  {B.}~\bibnamefont {Tong}}, \bibinfo {author} {\bibfnamefont {G.}~\bibnamefont
  {Bian}}, \bibinfo {author} {\bibfnamefont {N.}~\bibnamefont {Alidoust}},
  \bibinfo {author} {\bibfnamefont {C.-C.}\ \bibnamefont {Lee}}, \bibinfo
  {author} {\bibfnamefont {S.-M.}\ \bibnamefont {Huang}}, \bibinfo {author}
  {\bibfnamefont {T.-R.}\ \bibnamefont {Chang}}, \bibinfo {author}
  {\bibfnamefont {G.}~\bibnamefont {Chang}}, \bibinfo {author} {\bibfnamefont
  {C.-H.}\ \bibnamefont {Hsu}}, \bibinfo {author} {\bibfnamefont {H.-T.}\
  \bibnamefont {Jeng}}, \bibinfo {author} {\bibfnamefont {M.}~\bibnamefont
  {Neupane}}, \bibinfo {author} {\bibfnamefont {D.~S.}\ \bibnamefont
  {Sanchez}}, \bibinfo {author} {\bibfnamefont {H.}~\bibnamefont {Zheng}},
  \bibinfo {author} {\bibfnamefont {J.}~\bibnamefont {Wang}}, \bibinfo {author}
  {\bibfnamefont {H.}~\bibnamefont {Lin}}, \bibinfo {author} {\bibfnamefont
  {C.}~\bibnamefont {Zhang}}, \bibinfo {author} {\bibfnamefont {H.-Z.}\
  \bibnamefont {Lu}}, \bibinfo {author} {\bibfnamefont {S.-Q.}\ \bibnamefont
  {Shen}}, \bibinfo {author} {\bibfnamefont {T.}~\bibnamefont {Neupert}},
  \bibinfo {author} {\bibfnamefont {M.}~\bibnamefont {{Zahid Hasan}}}, \ and\
  \bibinfo {author} {\bibfnamefont {S.}~\bibnamefont {Jia}},\ }\href {\doibase
  10.1038/ncomms10735} {\bibfield  {journal} {\bibinfo  {journal} {Nature
  Communications}\ }\textbf {\bibinfo {volume} {7}},\ \bibinfo {pages} {10735}
  (\bibinfo {year} {2016})},\ \Eprint {http://arxiv.org/abs/1601.04208}
  {arXiv:1601.04208 [cond-mat.mtrl-sci]} \BibitemShut {NoStop}%
\end{thebibliography}%
\end{document}